\documentclass[defaultstyle,12pt]{article}

\usepackage{latexsym} 
\usepackage{graphicx}       
\usepackage{rotating}      
\usepackage{amssymb}
\usepackage{amsmath} 

\begin{document}

\title{Deterministic chaos in an ytterbium-doped mode-locked fiber laser}

\author{Lucas B. A. M\'{e}lo, Guillermo F. R. Palacios, Pedro V. Carelli, \\
L\'{u}cio H. Acioli, Jos\'{e} R. Rios Leite and Marcio H. G. de Miranda\\
\textit{Departamento de F\'{i}sica, Universidade Federal de Pernambuco,}\\ 
\textit{50740-560 Recife, PE - Brazil}\\
\textit{mhgm@df.ufpe.br}\\
Published in Optics Express, \textbf{26} 10, 13686 (2018).}

\maketitle

\begin{abstract} We experimentally study the nonlinear dynamics of a femtosecond ytterbium doped mode-locked fiber laser. With the laser operating in the pulsed regime a route to chaos is presented, starting from stable mode-locking, period two, period four, chaos and period three regimes. Return maps and bifurcation diagrams were extracted from time series for each regime. The analysis of the time series with the laser operating in the quasi mode-locked regime presents deterministic chaos described by an unidimensional R\"{o}ssler map. A positive Lyapunov exponent $\lambda = 0.14$ confirms the deterministic chaos of the system. We suggest an explanation about the observed map by relating gain saturation and intra-cavity loss. 
\end{abstract}

\section{Introduction}

The possibility of generating high-energy ultrashort pulses from Ytterbium Doped Mode-Locked Fiber Lasers (YDMLFL) has driven the interest of several groups due to a multitude of applications in science, medicine and industry~\cite{Hartl, Grelu, Wise}. As these lasers emit light at approximately 1000 nm where the group velocity dispersion (GVD) of most fibers is normal ($\beta_{2} >0$), they operate in a regime which is different from that of most of the erbium based femtosecond fiber lasers. It is well established that the intracavity behavior of the pulses in Er and Yb fiber lasers are quite different. In a solitonic Er laser~\cite{Matsas, Kelly} the pulse changes per round trip are relatively small compared to what occurs in an Yb based positive fiber dispersion with gratings for GVD compensation, or in stretched-pulse erbium fiber lasers~\cite{Tamura}, for which major pulse changes are usually present. This difference affects the self-phase modulation and self-amplitude modulation effects and, in principle, could lead to a completely different chaotic behavior. 

In the majority of the applications of mode-locked lasers the main interest is to obtain a stable periodic pulse train source, with equal intensity pulses. Mode-locked operation is achieved by properly chosen parameters for the laser cavity and gain media, searching for optimum stability in optical frequency, pulse intensity and repetition rate. On the other hand, by manipulating pump power, dispersion and intra-cavity power this scenario may change radically, where the pulse intensities may reveal a chaotic distribution.

Chaos in femtosecond lasers has long been a subject of interest~\cite{Luo}. For mode-locked titanium-sapphire lasers, for example, it is possible to obtain an apparently random distribution of pulse intensities at a fixed repetition frequency, which actually corresponds to a well studied chaotic regime~\cite{Hnilo, Hnilo2, Hnilo3}. For an Er fiber mode-locked lasers, Zhao et. al. studied nonlinear dynamics~\cite{Zhao}, where they experimentally observed dynamical regimes by recording time series and performed numerical studies of the dynamics using coupled Ginzburg-Landau equation.

However, to the best of our knowledge, a proper experimental characterization of the nonlinear dynamics of  pulsed Yb fiber lasers is lacking. One of the possible reasons for this is that the parameter space for this laser is larger than that of the erbium laser: in the YDMLFL one has the extra freedom to change the total dispersion, aside from pump power and polarization settings. This makes it difficult to systematically change the parameters and select specific dynamical regimes~\cite{Sucha, Morgner, Mozdy, Smi}. Most of the work done in nonlinear dynamics for Yb is theoretical~\cite{San, Hab, Tok, Kut, Yang}, and some experimental work present different types of phenomena such as rogue waves~\cite{Runge, Liu}, noise-like pulse formation~\cite{Ko, Za, Su}, soliton explosion~\cite{Erk} and soliton molecule formation~\cite{Grelu}. As pointed out by references~\cite{San, Qin}, YDMLFL constitutes a rich nonlinear system, used in several different applications. This leads us to conclude of the importance of a proper experimental characterization of unusual operating regimes.

In this work we present experimental scenarios of YDMLFL dynamics tuned from regular mode-locked regime to deterministic pulsed chaotic behavior. These regimes are characterized by using pump power as a control parameter of the nonlinear dynamics. In contrast with reference~\cite{Zhao}, we present bifurcation diagrams and return maps extracted from time series, showing clear distinction among five different dynamical regimes, also captured in the radio-frequency (RF) spectrum. We give experimental evidence that the laser has deterministic chaos by showing a functional relation between neighboring pulses, calculating the Lyapunov exponent for the chaotic dynamical regime and comparing with an unidimensional map behavior. We propose an explanation to the observed behavior which relates gain saturation and intra-cavity losses.

\section{Experiment}

\begin{figure}[htbp]
\centering
\includegraphics[width=0.9 \linewidth]{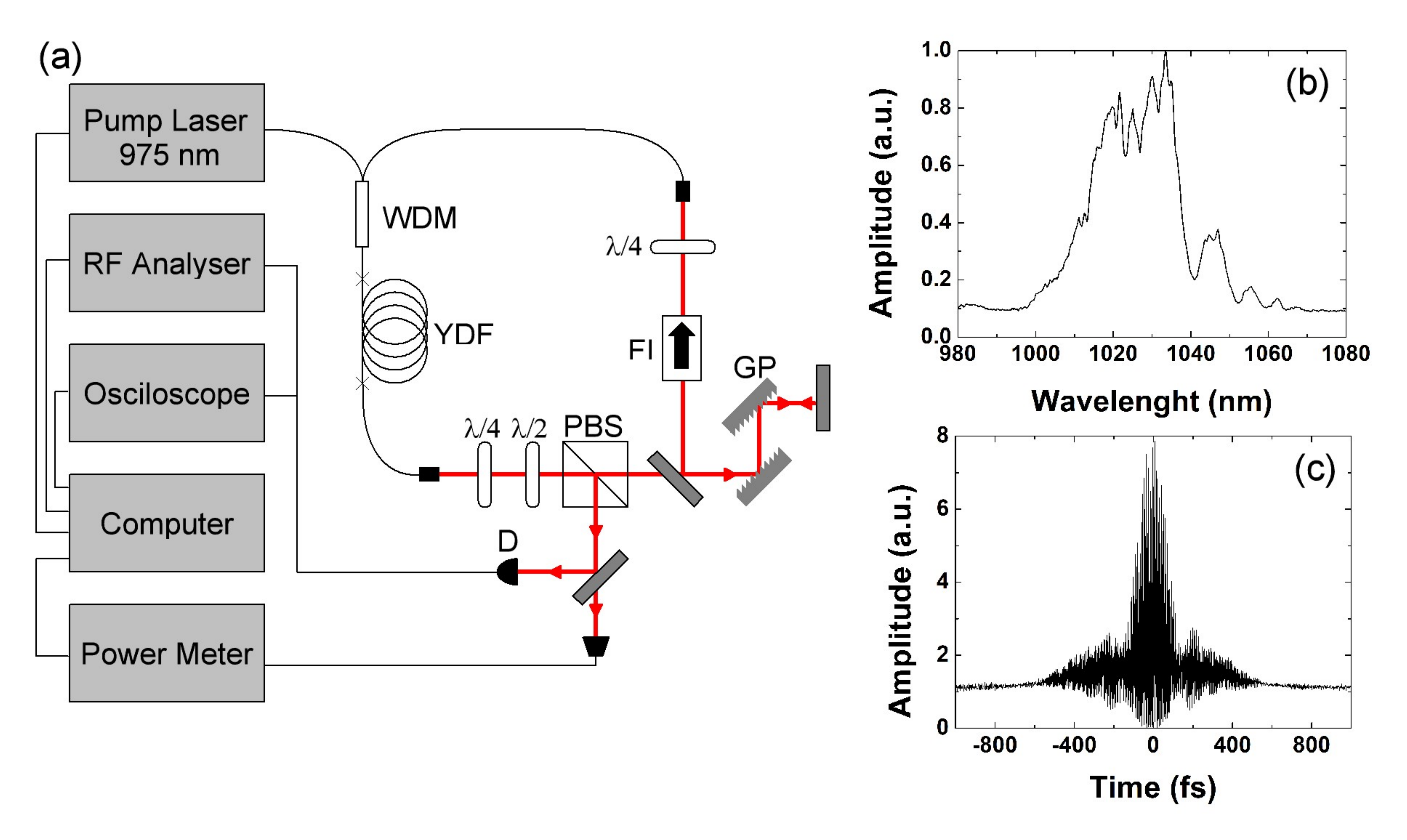}
\caption{(a) Schematics of the YFMLFL and data acquisition system: ytterbium doped fiber (YDF), Faraday isolator (FI), grating pair (GP), polarization beam splitter (PBS), wavelength division multiplexer (WDM) and detector (D). (b) Typical optical spectrum and (c) autocorrelation when the laser is in mode-locked operation. The optical spectrum and the autocorrelation were taken directly from the laser output, after the PBS.}
\label{expset}
\end{figure}

The experimental setup, shown in Fig.~\ref{expset}(a), consists of a home-made mode-locked Yb fiber laser. The pump laser operates at $\lambda_{pump} = 975$ nm. The gain medium is a single mode Yd doped fiber with an absorption coefficient of 1348 dB/m at 975 nm and length $L \approx 22$ cm. A set of waveplates (two $\lambda/4$ and one $\lambda/2$) and a polarizing beam splitter (PBS) are used to control the input and output polarization to the fiber section of the laser, the nonlinear polarization rotation in the fiber and therefore the nonlinear amplitude modulation of the pulses~\cite{Hof}. The grating pair (600 lines/mm) controls the intra-cavity dispersion of the pulse. An optical isolator guarantees the unidirectionality of the pulse circulation. The repetition frequency is $f_{rep} \approx 130$ MHz and optical spectrum centered at 1025 nm as shown in Fig.~\ref{expset}(b). The typical average power is $\left\langle P\right\rangle = 80$ mW and the pulse temporal duration is $\tau_p \approx 140$ fs, as measured from the colinear autocorrelation shown in Fig.~\ref{expset}(c). 

The data acquisition system involves a home-made photo-detector (1 GHz bandwidth), an oscilloscope (1 GHz bandwidth), a RF spectrum analyzer and a power meter, all computer-controlled as shown in Fig.~\ref{expset}(a). The diode pump laser current, $I_{pump}$, was scanned and the laser pulse train or time series was captured. Time series are acquired by the oscilloscope and saved as 100 ns windows (13 pulses). For each of the current settings of the pump laser, 10 windows are acquired and processed. The data acquisition sequence is: first the time series, then the RF spectrum and finally the average power are registered for a given $I_{pump}$. After one sequence, $I_{pump}$ is incremented by 1 mA and another sequence is taken. The procedure was repeated for a current range of 560 to 720 mA for a total of 160 sequences. The processing program maps the pulses peak values from time series to create bifurcation diagrams as function of $I_{pump}$ and return maps for the pulse peak values. As the detector response is much slower than the femtosecond pulses, the peak values are proportional to the pulse energy. The periodic and chaotic nature of the dynamics is still faithfully represented by the data.

\section{Results and discussion}

\begin{figure}[htbp]
\centering
\includegraphics[width=0.9\linewidth]{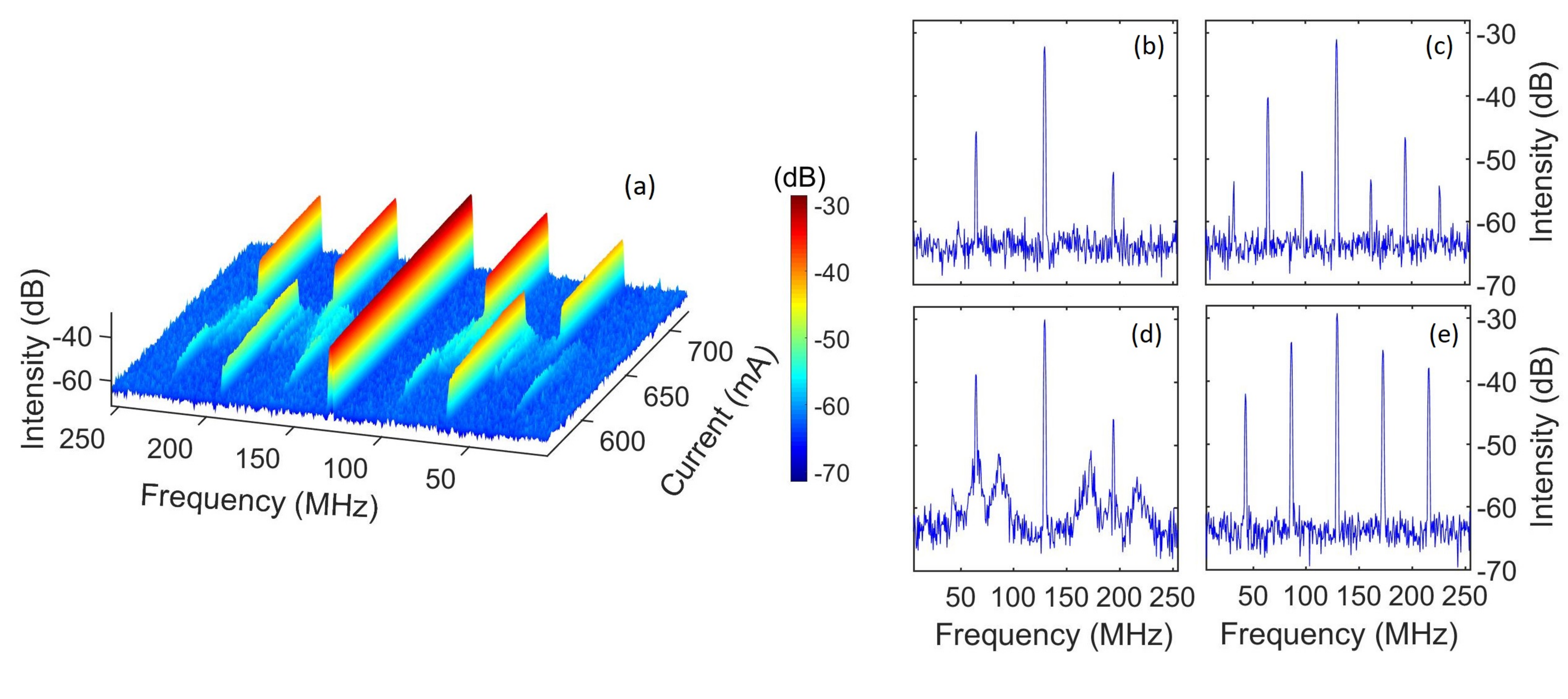}
\caption{(a) 3D Radio Frequency spectrum showing period two, four, chaos and period three. The repetition frequency is $f_{rep} \approx 130$ MHz. RF spectra for individual currents for period two, four, chaos and three are presented in (b), (c), (d) and (e) respectively.}
\label{rfs}
\end{figure}

The first recognizable feature of chaotic dynamics appears in the evolution of the RF spectrum of the pulse train as $I_{pump}$ is increased. This behavior is systematically registered in Fig.~\ref{rfs}(a) which shows a 3D graph of 160 RF spectra as a function of $I_{pump}$. The RF spectra for frequencies greater than $f_{rep}$ are higher order beat notes of the peaks below $f_{rep}$, and are simply mirror of the later. For $560~\text{mA} < I_{pump} < 566~\text{mA}$ only a single peak is seen at $f_{rep} \approx 130$ MHz. The laser operates in a stable mode-locked regime and the pulse train has a constant peak amplitude. For $566~\text{mA} < I_{pump} < 584~\text{mA}$ a new peak at $f_{rep}/2 \approx 65$ MHz becomes present, corresponding to period doubling. For currents $585~\text{mA} < I_{pump} < 610~\text{mA}$, another peak appears at a frequency $f_{rep}/4 \approx 32.5$ MHz, corresponding to period quadrupling. When $610~\text{mA} < I_{pump} < 652~\text{mA}$ there is a dramatic change: the RF spectra presents a multitude of broad peaks typical of chaotic behavior. Increasing the current even further, for $653~\text{mA} < I_{pump} < 721~\text{mA}$ a new peak appears at $f_{rep}/3 \approx 43$ MHz. RF spectra for the currents for period two, four, chaos and three are presented in Figs.~\ref{rfs}(b), \ref{rfs}(c), \ref{rfs}(d) and \ref{rfs}(e) respectively.

\begin{figure}[htbp]
\centering
\hspace{0.0mm}\includegraphics[width=1.0\linewidth]{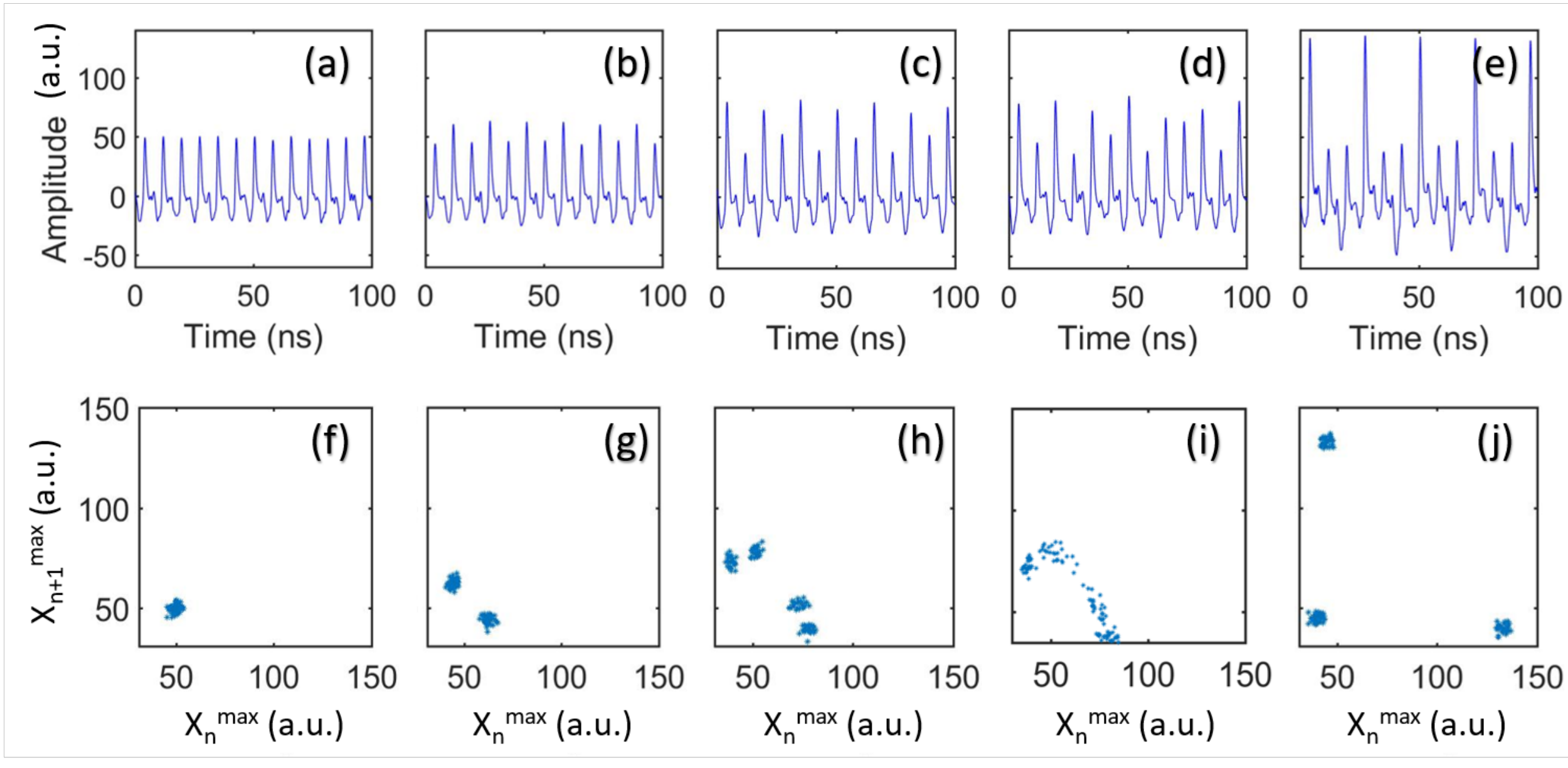}
\caption{Time series (upper row) of the laser pulses and the corresponding peak value return maps (RM) (lower row): in (a) the laser is stable and the RM (f) has only one spot. In (b) the period two is seen and the RM (g) shows two distinct spots. In (c) period four dynamics is seen, after the second period doubling bifurcation and the RM (h) has four distinct spots. In (d), the pulse intensity is chaotic and the RM (i) is scattered over a relatively well defined region indicating deterministic chaos describable by a unidimensional map. In (e) there is a stable period three regime and the RM in (j) shows three dots.}
\label{ts}
\end{figure}

The nonlinear dynamics observed in the RF spectra is also revealed by the time series of the pulse train as $I_{pump}$ is scanned. First the laser output presents a stable periodic pulse train, as shown in Fig.~\ref{ts}(a) with constant pulse energy. Next we observe a period two regime after the first period doubling bifurcation, as shown in Fig.~\ref{ts}(b). Further increasing the pump power produces a period four regime after a second period doubling bifurcation, as seen in Fig.~\ref{ts}(c). These bifurcations indicate the beginning of a period doubling cascade. However, due to scanning and detection resolution was not possible to observe higher order like period eight. In continuation, Fig.~\ref{ts}(d) presents chaotic behavior, when the pulse energies assume apparently random values. Continuing to increase $I_{pump}$ results in the appearance of the stable period three regime shown in Fig.~\ref{ts}(e) that emerges as in the saddle-node bifurcation for stable period three in unidimensional maps like the logistic map.

From the experimental time series we obtain the return maps (RM) plotting the peak value $x^{max}_{n+1}$ versus the previous peak value $x^{max}_{n}$. Fig.~\ref{ts}(f) shows the RM when the laser is stable and has a single peak power, corresponding to Fig.~\ref{ts}(a). Figs.~\ref{ts}(g) and \ref{ts}(h) show the RMs for the period two and period four regimes, corresponding to Figs.~\ref{ts}(b) and \ref{ts}(c), respectively. Fig.~\ref{ts}(i) shows the RM when the regime is chaotic. In this case, the points are distributed close to a parabola-like curve, displaying a deterministic characteristic relation between $x^{max}_{n+1}$ and $x^{max}_{n}$, typical of highly dissipative systems that have unidimensional and unimodal map. Finallly, Fig.~\ref{ts}(j) shows that the RM presents a period three bifurcation corresponding to Fig.~\ref{ts}(e).

\begin{figure}[htbp]
\centering
\includegraphics[width=0.6\linewidth]{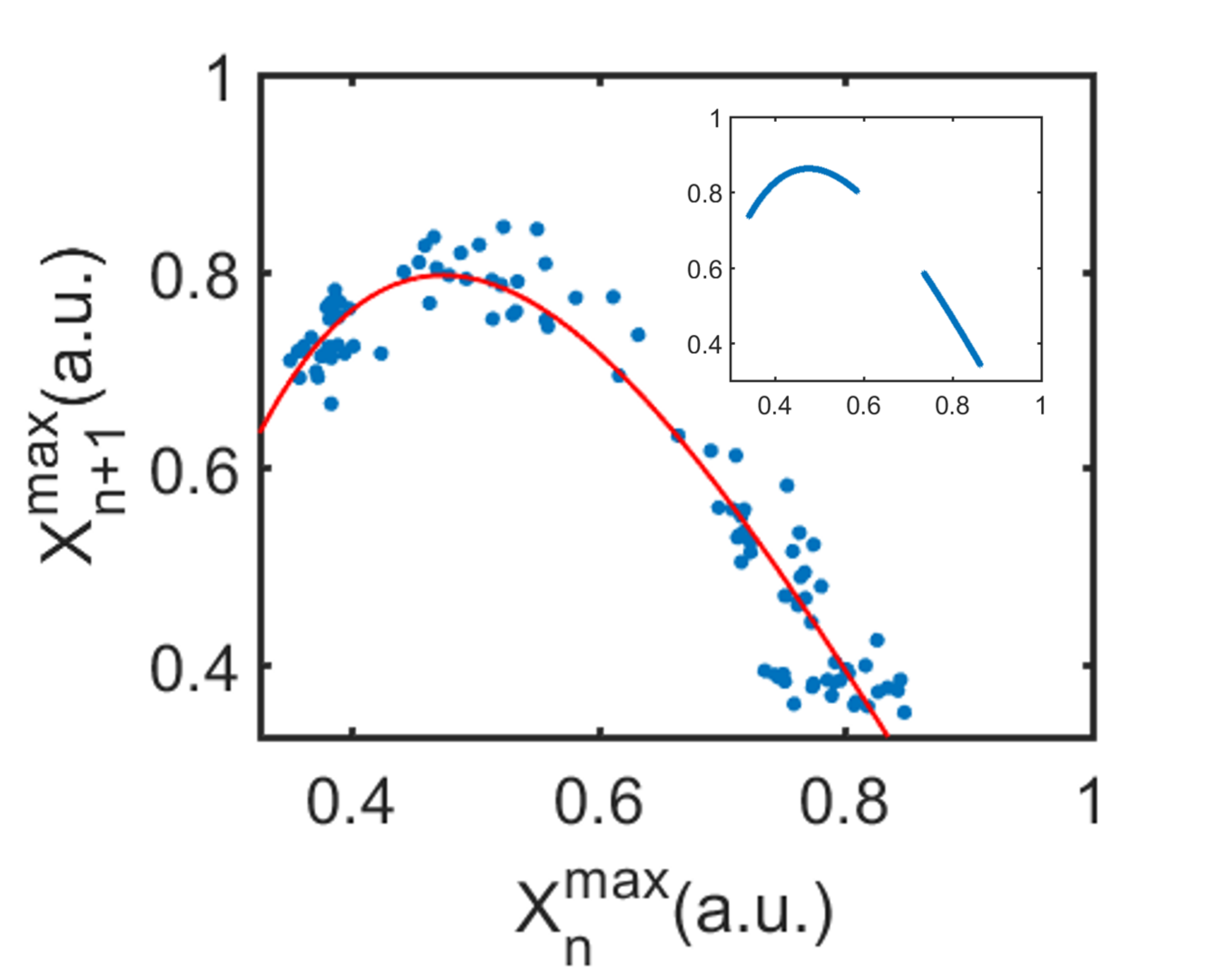}
\caption{Return map at the chaotic region from Fig.~\ref{ts}(i). This experimental map was fitted (continuous curve, red color) using a R\"{o}ssler map. The inset figure is a set of points obtained by iterating the R\"{o}ssler map function.}
\label{fit}
\end{figure}

In order to provide a quantitative characterization of the operation mode of our laser we have calculated the Lyapunov exponent, $\lambda$, for the data presented in Fig.~\ref{ts}(i), which is plotted and rescaled in Fig.~\ref{fit}, for better visualization. For a unidimensional map $\lambda$ is given by~\cite{Strogatz},
\begin{equation}
\lambda =  \lim_{N \rightarrow \infty} \frac{1}{N}\sum_{n=1}^{N}\ln\left| \frac{df(x_{n})}{dx_n}\right|.
\label{eq1}
\end{equation}
The Lyapunov exponent was calculated using the experimental data series points and also a support function $f(x)$ adjusted to the data, as shown in Fig.~\ref{fit}. We used $f(x)$ to calculate the derivative at each point, since there were only 100 data points. This function was chosen based on the form of the experimental data in Fig.~\ref{fit} and its resemblance to a R\"{o}ssler map~\cite{Rox}. Assuming $f(x) = Ax\exp(Bx)+C$, we obtain $A = 15.2$, $B = -2.1$ and $C = -1.8$ as best fit parameters. The resulting fit (red line) is shown at Fig.~\ref{fit} superimposed on the experimental data. The Lyapunov exponent obtained from the adjusted $f(x)$ is $\lambda = 0.14$. The positive value indicates that the dynamic regime is chaotic. The inset curve in Fig.~\ref{fit} is a set of points in the chaotic regime reconstructed numerically from $f(x)$ with the fit parameters $A$, $B$ and $C$. This curve shows a good agreement with the experimental data.

To understand the origin of the nonlinear behavior observed in our laser we explain its operation based on the following ingredients: first, nonlinear losses due to nonlinear polarization rotation and second, pulse energy dependent gain saturation~\cite{San}. The first contribution is instantaneous and reduces the losses for high energy pulses, acting directly on each pulse. Gain saturation, on the other hand, acts on the subsequent pulses of the pulse train: the passage of a high energy pulse means that the gain medium does not recover completely for the next pulse which therefore experiences a smaller gain. In the perfectly mode-locked regime the gain is periodically modulated with constant amplitude and the system is temporally invariant. This means there is a fixed point in a phase space description. A simple model which leads to a nice geometrical interpretation of the laser dynamics, based solely on gain saturation and saturation of the nonlinear loss, has been proposed by Li et al~\cite{Feng}, and later expanded by Wei et al~\cite{Hu}. From the model presented in~\cite{Feng} a picture similar to the logistic map is obtained and a period doubling route to chaos using pulse saturation energy as a control parameter is clearly demonstrated numerically. Interestingly, Li et al~\cite{Feng} also predict the existence of multiple pulsing behavior after the chaotic regime, which we have not observed in our laser, but as pointed in~\cite{Feng}, the details of the nonlinear loss curve as a function of pulse energy is critical to determine the laser's dynamics. We believe that the differences in the dispersion regime of the Yb laser compared to the erbium soliton laser, for example, are sufficient to allow reaching different dynamic regimes. 

\begin{figure}[ht]
\centering
\begin{tabular}{cc}
\hspace{-6.0cm}(a)&\hspace{-6.0cm}(b)\\
\hspace{0.0cm}\includegraphics[width=0.5\linewidth]{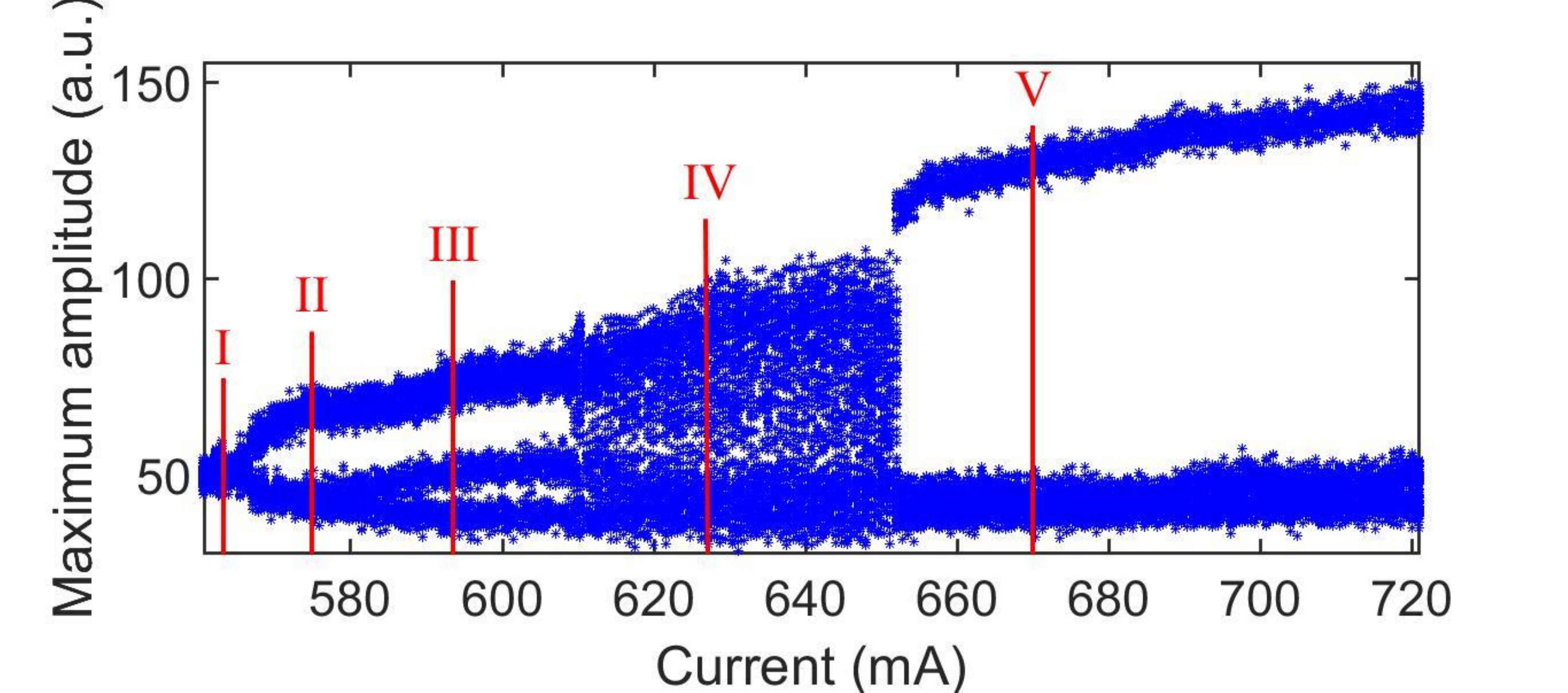}&\hspace{0.0cm}\includegraphics[width=0.5\linewidth]{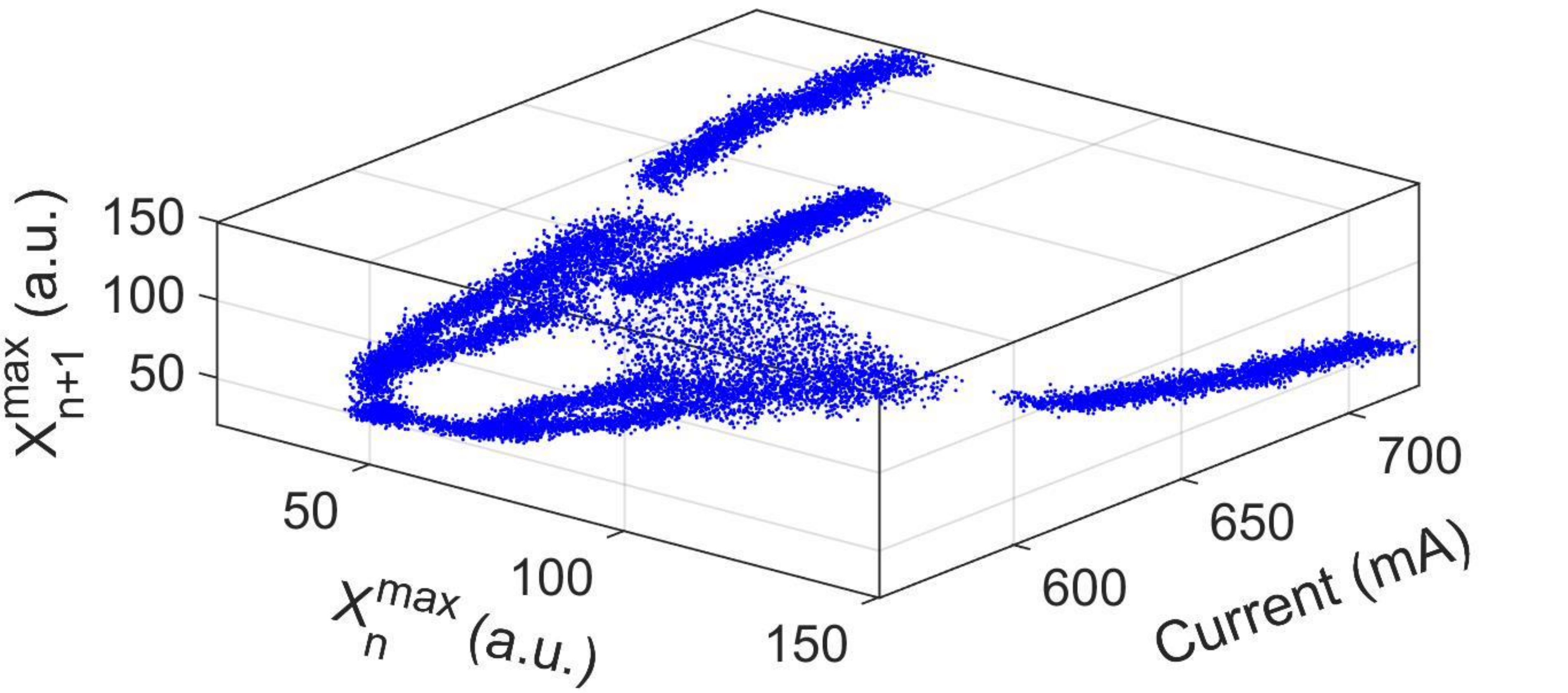}
\end{tabular}
\caption{Bifurcation diagrams for Yb mode-locked laser: (\textbf{a}) is a conventional bifurcation diagram. The five red lines show five different dynamical regimes related to the time series of Fig.~\ref{ts}: (I) the laser is at standard mode-locked operation, (II) period two, (III) period four, (IV) chaos and (V) period three window. (\textbf{b}) is a tridimensional bifurcation diagram for a better visualization of all branches. In the supplementary material there is an animation of this figure going from 2D to 3D bifurcation diagram.}
\label{bd}
\end{figure}

Experimental bifurcation diagram of the time series was also used to visualize the dynamical scenarios. The importance of bifurcation diagrams lies on the fact that they can indicate the route towards the chaotic regime as the control parameter, $I_{pump}$, is changed. Fig.~\ref{bd}(a) shows the experimental bifurcation diagram with the positions of the five RMs already described in Fig.~\ref{ts}. The RMs are transversal slices taken from Fig.~\ref{bd}(a). The RM of Fig.~\ref{ts}(f) represents a stable pulse train, therefore there is only one branch at the bifurcation diagram. The RM from Fig.~\ref{ts}(g) have the first bifurcation, so there are two branches. The RM from Fig.~\ref{ts}(h) has a second bifurcation but there are only three branches visible because the other one is hidden by the noise blur. This branch is visualized in the 3D bifurcation diagram shown in Fig.~\ref{bd}(b), where two consecutive values of the pulse peaks are depicted. The RM from Fig.~\ref{ts}(i) shows the chaotic regime. The RM from Fig.~\ref{ts}(j) that presents a bifurcation with three branches, also has one branch hidden in Fig.~\ref{bd}(a), which is visible in Fig.~\ref{bd}(b).

\section{Conclusion}

In conclusion, following a systematic experimental procedure, we show that Yb mode-locked fiber laser has striking features of low dimension deterministic chaos. We have characterized the nonlinear dynamical regimes by measuring the RF spectrum and return maps of mode-locked pulses in a range of pump laser current where there are period doubling, period tripling and quadrupling of pulse peak intensity. We obtain conventional and 3D bifurcation diagrams, showing characteristics of a period doubling cascade route to chaos with periodic windows as in unimodal maps like the logistic and the R\"{o}ssler maps. Return maps for the next pulse peak were obtained and the R\"{o}ssler map was used to fit the chaotic regime. This function was used to reconstruct numerically the chaotic regime, showing a good agreement with the experimental data. Using this function, the Lyapunov exponent was obtained with positive value, corroborating that this is a very dissipative system with similar features of a R\"{o}ssler map. Hopefully, our results will impact theoretical models for mode-locked Yb fiber laser dynamics.

\section*{Funding} 

Conselho Nacional de Desenvolvimento Cient\'{i}fico e Tecnol\'{o}gico (CNPq) ($441668/2014-3$); Funda\c{c}\~{a}o de Amparo \`{a} Ci\^{e}ncia e Tecnologia do Estado de Pernambuco (FACEPE) (APQ$-1178-10.5/14$, APQ$-0826-10.5/15$).


\end{document}